\documentclass[%
 reprint,
 amsmath,amssymb,
 aps,
]{revtex4-1}

\usepackage{color}
\usepackage{graphicx}
\usepackage{dcolumn}
\usepackage{bm}

\begin{document}

\preprint{APS/123-QED}

\title{Band-gap and Band-edge Engineering of Multicomponent Garnet Scintillators: A First-principles Study }

\author{S.K. Yadav}
 \email{syadav@lanl.gov, yadav.satyesh@gmail.com}
\author{B.P. Uberuaga}%
\affiliation{%
Materials Science and Technology Division, Los Alamos National Laboratory, Los Alamos, NM 87545, USA}%


\author{M. Nikl}
\affiliation{
Institute of Physics, Academy of Sciences of the Czech Republic, 16253 Prague, Czech Republic}%

\author{C. Jiang}
\affiliation{%
Thermo-Calc Software Inc., Pittsburgh, PA 15317, USA}%

\author{C.R. Stanek}
\affiliation{%
Materials Science and Technology Division, Los Alamos National Laboratory, Los Alamos, NM 87545,USA }%

\author{LA-UR-15-23403}


\date{\today}

\begin{abstract}
Complex doping schemes in RE$_3$Al$_5$O$_{12}$ (RE=rare earth element) garnet compounds have recently led to pronounced improvements in scintillator performance.  Specifically, by admixing lutetium and yttrium aluminate garnets with gallium and gadolinium, the band-gap was altered in a manner that facilitated the removal of deleterious electron trapping associated with cation antisite defects.  Here, we expand upon this initial work to systematically investigate the effect of substitutional admixing on the energy levels of band edges. Density functional theory was used to survey potential admixing candidates that modify either the conduction band minimum (CBM) or valence band maximum (VBM). We considered two sets of compositions based on Lu$_3$B$_5$O$_{12}$ where B = Al, Ga, In, As, and Sb; and RE$_3$Al$_5$O$_{12}$, where RE = Lu, Gd, Dy, and Er.  We found that admixing with various RE cations does not appreciably effect the band gap or band edges. In contrast, substituting Al with cations of dissimilar ionic radii has a profound impact on the band structure. We further show that certain dopants can be used to selectively modify only the CBM or the VBM. Specifically, Ga and In decrease the band gap by lowering the CBM, while As and Sb decrease the band gap by raising the VBM. These results demonstrate a powerful approach to quickly screen the impact of dopants on the electronic structure of scintillator compounds, identifying those dopants which alter the band edges in very specific ways to eliminate both electron and hole traps responsible for performance limitations. This approach should be broadly applicable for the optimization of electronic and optical performance for a wide range of compounds by tuning the VBM and CBM.
\end{abstract}

\pacs{Valid PACS appear here}
\maketitle


\section{\label{sec:intro}Introduction}

A$_3$B$_5$O$_{12}$ garnets, and in particular RE$_3$Al$_5$O$_{12}$ compositions (where RE is rare earth element or Y), have been studied for technical use as optical materials for over 50 years \cite{geusic,blasse-b,weber-g}.  Although garnets also received interest as a scintillator $\sim$20 years ago \cite{moszynski-b}, a lower light yield than other compounds ultimately led to relative disinterest. Often, defects trap charge carriers otherwise available to participate in the scintillation process, thus potentially resulting in delayed and/or reduced light output. The important role that defects play in scintillator performance has been well-documented \cite{weber-f}. However, recent studies involving co-doping of garnets have demonstrated dramatic improvements in light yield and these findings have consequently reinvigorated interest in garnets as high performance scintillators \cite{nikl2008scintillator,kanai2008characteristics,cherepy2010transparent,kamada2011scintillator,kamada2011composition,nikl2013development,tyagi2013effect,ogieglo2013luminescence,drozdowski2014studies,wu2014role,sibczynski2015characterization}.  These optimization efforts have relied on the manipulation of the garnet electronic structure through admixing, and in the process creating so-called ``multicomponent" garnets \cite{nikl-review}.  It is well-known that cation antisite defects are present in garnets (RE$^{3+}$ on Al$^{3+}$ sites and vice versa) \cite{lupei-b,nikl-c,donnerberg,kuklja,chiara,stanek-i} and that they contribute to reduced scintillator performance \cite{nikl-d}  by creating traps for the electronic carriers which results in considerable slowing down of scintillation response.  However, the challenge of removing cation antisite defects in garnet is that they are isovalent (i.e. charge neutral) and the corresponding defect formation energy is rather low, thereby preventing a defect engineering approach.  Therefore alternative defect management methods are required.  Interestingly, it has been shown that adding Ga  to aluminate garnets removes the signature of antisite defects \cite{nikl-b}.  This implies that Ga admixing eliminates the effectiveness of the antisite traps. However, Ga is closer in size to the RE cation than it is to Al \cite{shannon}, which suggests that a higher concentration of antisites should exist in Ga-doped garnets than in pure aluminate garnets - a hypothesis validated by a joint experimental-atomistic simulation study \cite{shirinyan,stanek-o}.  Rather, instead of reducing the concentration of deleterious antisite defects, the benefit of Ga-admixing arises from shifts in the conduction band such that it envelops the trap state in the forbidden gap associated with the antisite defect \cite{stanek-p,nikl-review}.  This is a primary example of the ``band-gap engineering'' approach to defect management.

\begin{figure*}
\includegraphics[width=7in]{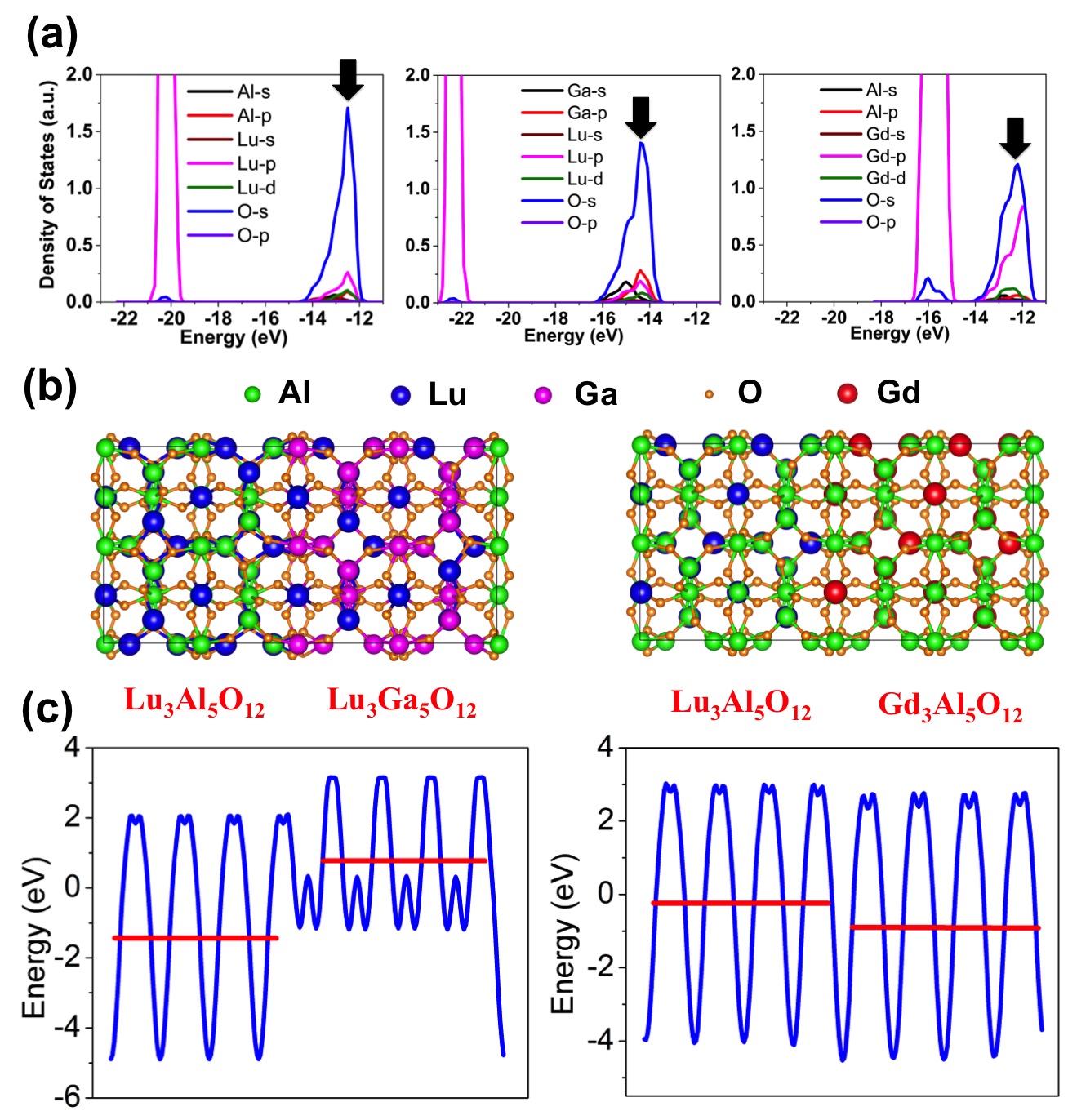}
\caption {(Color online) (a) Density of states of deep states of Lu$_3$Al$_5$O$_{12}$, Lu$_3$Ga$_5$O$_{12}$, and Gd$_3$Al$_5$O$_{12}$, where the oxygen state that is used as a reference is marked by an arrow. (b) Supercell used for calculating electrostatic potentials, where Lu is blue, Al green, Ga magenta, O orange, Gd red. (c) Electrostatic potential calculated using the supercell in (b), where the red horizontal line indicates the average electrostatic potential.}
{\label{alignment}}
\end{figure*}

In this paper, we build upon our previous effort to optimize the electronic structure of multicomponent garnets by studying a range of dopants and their effect on the energy levels of band-edges of Lu$_3$Al$_5$O$_{12}$. By using density functional theory (DFT) based first-principles calculations, we show that certain 3+ dopants that substitute for Al can result in variations in either the valence or conduction band edges, while leaving the other band edge more or less unchanged - thus opening the path for ``band-edge engineering" through admixing.  We also show that substituting Lu with RE cations does not have a significant impact on band edges.  Although we use garnet as a case study, it is anticipated that this approach can be extended to a wide range of scintillator compounds and provide an efficient manner to screen dopants for optimizing performance.

\section{\label{sec:method}Methodology}

\subsection{First-principles method}

Density functional theory (DFT) calculations were performed using the Vienna \textit{Ab initio} Simulation Package (VASP) \cite{kresse1996efficient}. The DFT calculations employed the Perdew, Burke, and Ernzerhof (PBE) \cite{perdew1996generalized} generalized gradient approximation (GGA) exchange-correlation functional and the projector-augmented wave (PAW) method \cite{blochl1994projector}. For all calculations, a plane wave cutoff of 500 eV for the plane wave expansion of the wave functions was used to obtain highly accurate forces. In the results reported, only the gamma point was considered in the k-space sampling; however, we have employed denser k-point meshes in select cases and very similar results were obtained. All structures were fully relaxed without any symmetry constraints and relaxations were considered converged when each component of the force on every atom was smaller than 0.02 eV/\AA.

\subsection{Band edge alignment}

\begin{table}
\begin{center}
\caption{Relative shift of band edges in eV as calculated using a deep $s$ state (Deep-state) and aligning the electrostatic potential (Electrostatic).}
\begin{tabular}{p{0.7in} p{.5in} p{0.7in} p{0.7in}} 
  \hline \hline
  
   { } &  & { Lu$_3$Ga$_5$O$_{12}$} & {Gd$_3$Al$_5$O$_{12}$}  \\
      {Method} & { } & {  } & {}  \\
\hline
{Deep-state} & {} & { 1.9} & {0.4}  \\
{Electrostatic} & {} & { 2.1} & {0.6}  \\
  \hline
  \hline
\end{tabular}
\label{table1}
\end{center}
\end{table}

The first step to reliably determine the relative position of band edges in a compound as a function of composition is to identify a reference state that does not change with chemical composition. There are several references that are used in the literature to determine the relative position of band edges \cite{wei1995band, moses2010band, van1986theoretical, franciosi1996heterojunction}. The average electrostatic potential is the best common reference but it is very expensive to calculate as both materials of interest must be contained within one common simulation cell.  Not only does this necessitate large cells to accommodate both materials, but also to avoid interfacial effects that are not of interest here. Rather than rely on computationally intensive electrostatic potential approach, in this work we use a deep $s$ state of oxygen as a reference to realign band edges of two compounds \cite{wei1995band}.

Figure ~\ref{alignment}(a) shows the density of states (DOS) plot of Lu$_3$Al$_5$O$_{12}$, Lu$_3$Ga$_5$O$_{12}$, and Lu$_3$Gd$_5$O$_{12}$, and the $s$ state chosen for comparison is indicated by an arrow. It can be seen in Figure ~\ref{alignment}(a) that the deep state chosen is dominant compared to all other orbitals at that energy making it an ideal candidate for band alignment as this state is insensitive to the local coordination of the atoms and thus should have the same energy regardless of environment. With this deep state identified, band edges of two systems then can be compared directly by shifting the band structure of one such that the energy of the deep state coincides with the same state in the other structure. 

To validate the approach of employing the computationally less intensive deep state approach, we compared the relative shift with the average electrostatic potential  for two cases (Al substitution with Ga, and Lu substitution with Gd). Fig.~\ref{alignment}(b) shows the supercell used for calculating the average electrostatic potential. The offset between the two systems was calculated using the average electrostatic potentials for Lu$_3$Al$_5$O$_{12}$ and Lu$_3$Ga$_5$O$_{12}$ and Lu$_3$Al$_5$O$_{12}$ and Gd$_3$Al$_5$O$_{12}$ . Table~\ref{table1} shows the good agreement between the two methods for calculating the band offset, providing confidence that the deep $s$ state approach gives physically meaningful values.   

\section{Results}
 
\subsection{Al substitution}

\begin{figure}
\includegraphics[width=3in]{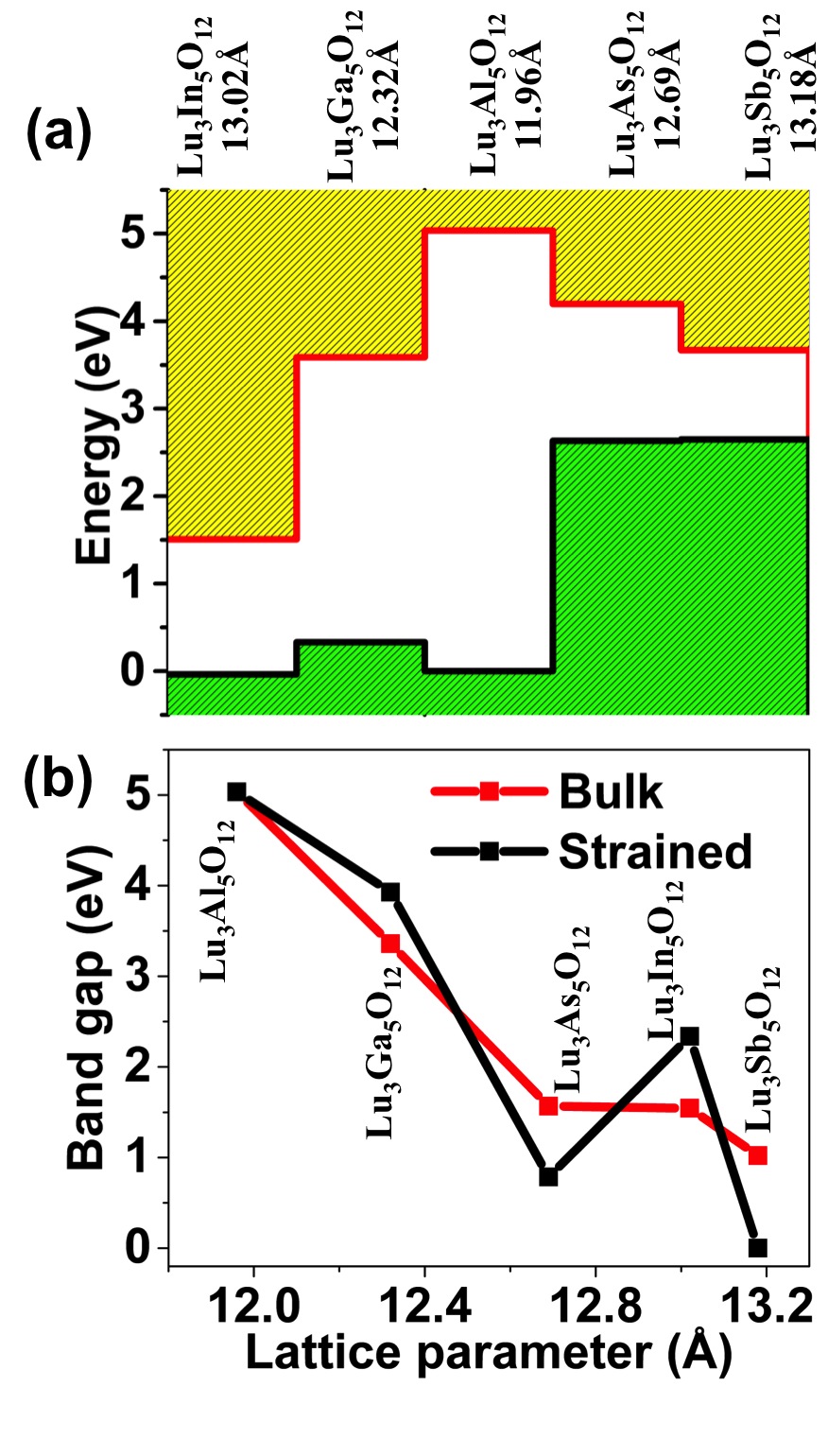}
\caption {(Color online) (a) Conduction and valence band change for Lu$_3$B$_5$O$_{12}$ compounds, where B = Al, In, Ga, As and Sb. (b) Change in band gap as a function of lattice parameter for the different compounds considered. The ``bulk'' (red) line indicates the band gap of the compound at its relaxed lattice constant while the ``strained'' (black) line is the band gap of the compound when placed at the lattice constant of Lu$_3$Al$_5$O$_{12}$, to better separate the roles of chemistry and strain on the changes in the band gap. The band gap for the strained compound is plotted versus the compound's natural lattice constant, for ease of comparison. }
{\label{al-substitution}}
\end{figure}

First we consider the extreme case of full Al substitution with larger cations. Using the deep $s$ state of oxygen as the common reference, we calculated the relative position of band edges of various Lu$_3$B$_5$O$_{12}$ compounds, where B=(Al, Ga, In, As, and Sb). Other garnets are less common than Al garnets, with only Lu$_3$Ga$_5$O$_{12}$ \cite{venkatramu2010nanocrystalline,xu2010cathodoluminescent,mahalingam2008bright} and Lu$_3$Sb$_5$O$_{12}$ \cite{raevsky1994nuclear} reported in literature. However, analyzing how other B cations impact the electronic structure may guide future doping and admixing strategies where full substitution may not be required. Figure~\ref{al-substitution}(a) shows the relative position of the conduction band minimum (CBM) and valence band maximum (VBM), along with the lattice parameters, for each of the compounds considered. Figure~\ref{al-substitution}(b) shows the change in band gap as a function of lattice parameter. There are several observations that can be made from from Fig.~\ref{al-substitution}.  First, Lu$_3$Al$_5$O$_{12}$ (LuAG) has the largest band gap of all compounds considered, and the band gap decreases with increasing lattice parameter.  As has been observed previously, for Lu$_3$Ga$_5$O$_{12}$ (LGG) the CBM is shifted with respect to LuAG, while the VBM is only slightly shifted and this shift in the CBM of LGG is related to the CBM shift observed in Ga-doped LuAG, which leads to the overlap of the cation antisite trap state~\cite{stanek-p}.  A similar, but more pronounced, effect is observed for Lu$_3$In$_5$O$_{12}$, where the CBM is further shifted with respect to LuAG and LGG, but the VBM remains near to that of LuAG and LGG.  Overall, while the VBM remains essentially constant when substituting Al with In and Ga, large CBM variations are observed. However, substituting Al with As and Sb leads to significantly larger variations in the VBM while the associated shifts in the CBM are relatively modest, see Figure ~\ref{al-substitution}. Thus, upon substitution of Al with As or Sb, the overall decrease in the band gap is primarily due to increased VBM energy.  Although the VBM shifts observed for Lu$_3$As$_5$O$_{12}$ and Lu$_3$Sb$_5$O$_{12}$ are similar, the larger Lu$_3$Sb$_5$O$_{12}$ exhibits a larger CBM shift.

\begin{figure}
\includegraphics[width=2.5in]{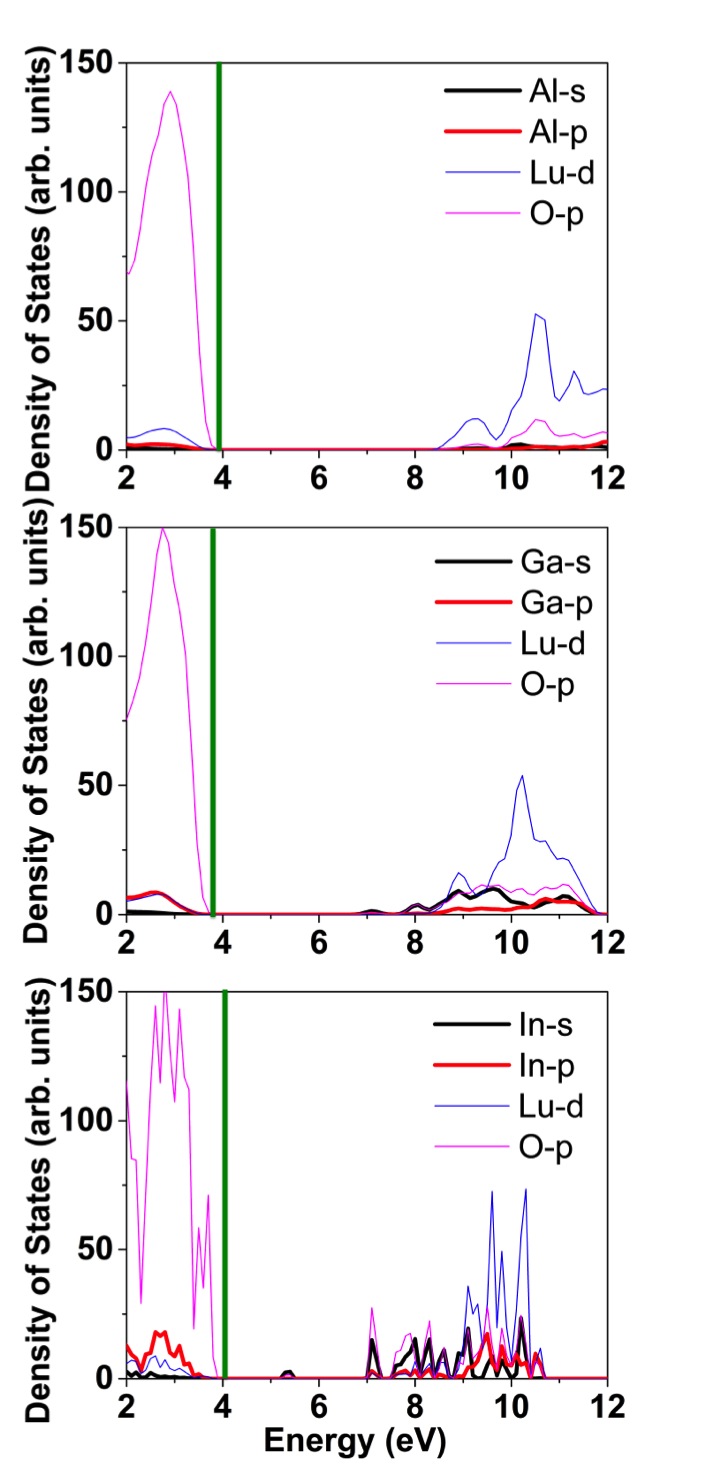}
\caption {(Color online)  Density of states arising from Al, Ga, In \textit{s} and \textit{p} states, Lu \textit{d} states and O \textit{p} states for garnets with B=Al, Ga, and In. The green vertical line corresponds to the Fermi level (highest occupied state), obtained by the alignment of the deep oxygen s state.}
{\label{io-dos}}
\end{figure}

\begin{figure}
\includegraphics[width=2.5in]{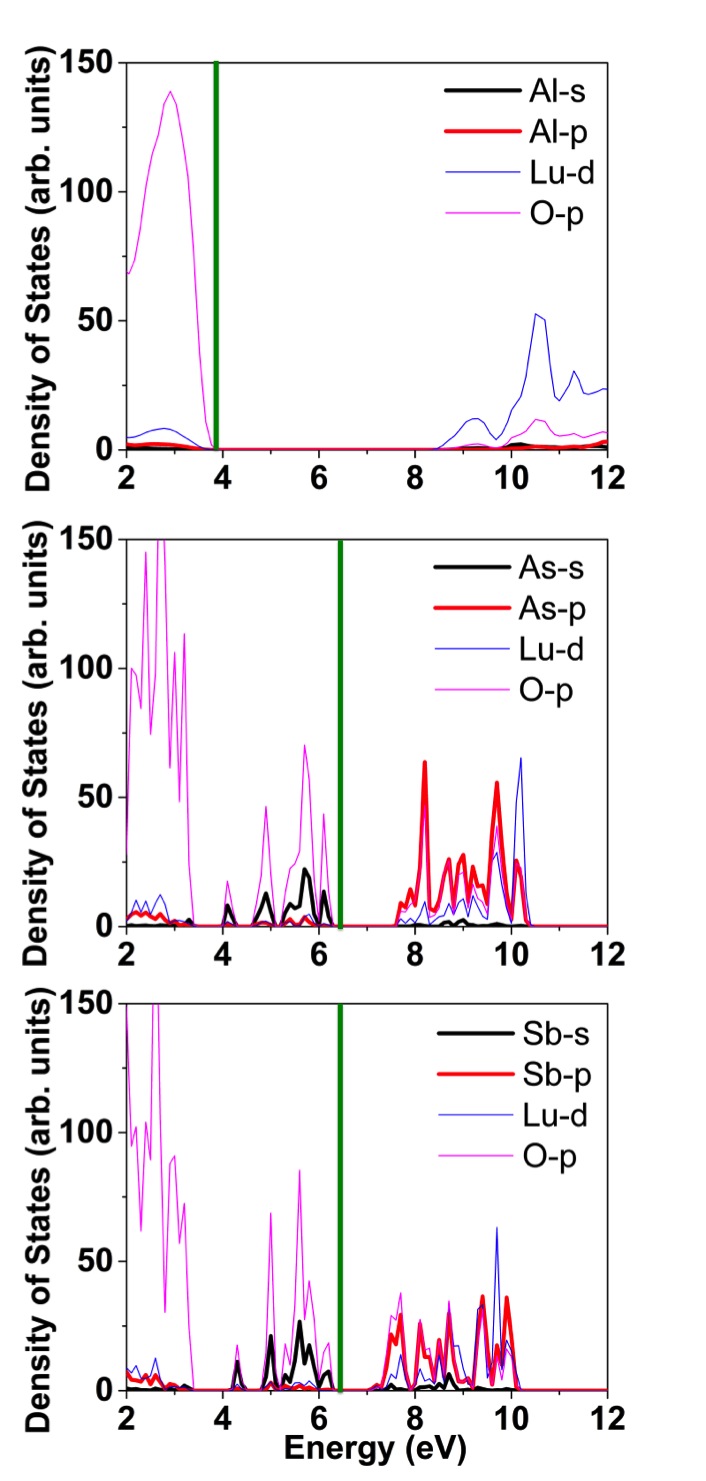}
\caption {(Color online) Density of states arising from Al, As, Sb \textit{s} and \textit{p} states, Lu \textit{d} states and O \textit{p} states for garnets with B=Al, As, and Sb. The green vertical line corresponds to the Fermi level (highest occupied state), obtained by the alignment of the deep oxygen s state.}
{\label{co-dos}}
\end{figure}

The difference between Lu$_3$Sb$_5$O$_{12}$/Lu$_3$In$_5$O$_{12}$ (i.e large VBM shift) and Lu$_3$In$_5$O$_{12}$/Lu$_3$Ga$_5$O$_{12}$ (i.e. large CBM shift) can be understood by closely examining the states that constitute the CBM and VBM. For example, Fig.~\ref{io-dos} shows the electronic density of states (DOS) for Lu$_3$Al$_5$O$_{12}$, Lu$_3$Ga$_5$O$_{12}$, and Lu$_3$In$_5$O$_{12}$. The DOS of Lu$_3$Al$_5$O$_{12}$ shows that the CBM is comprised of a Lu \textit{d} state and the VBM is dominated by an O \textit{p} state. In Lu$_3$Ga$_5$O$_{12}$ and Lu$_3$In$_5$O$_{12}$, the CBM shift is driven by Ga and In \textit{s} states which are, in these two cases, dominant contributors to the CBM.  Figure~\ref{co-dos} shows the DOS for Lu$_3$Al$_5$O$_{12}$, Lu$_3$As$_5$O$_{12}$, and Lu$_3$Sb$_5$O$_{12}$. In Lu$_3$As$_5$O$_{12}$ and Lu$_3$Sb$_5$O$_{12}$ the CBM shift is driven by As and Sb \textit{s} states which are now dominant contributors to the VBM along with the O \textit{p} state. In addition to the prominent difference in hybridization of states in the two cases, Bader charge analysis shows that As and Sb bonds are more covalent compared to Ga and In bonds~\cite{bader1990atoms,tang2009grid}. Thus, and as expected, more pronounced covalent bonding pushes the VBM higher in energy while ionic bonding shifts the CBM down in energy to reduce the overall band gap. 

\begin{figure}
\includegraphics[width=3in]{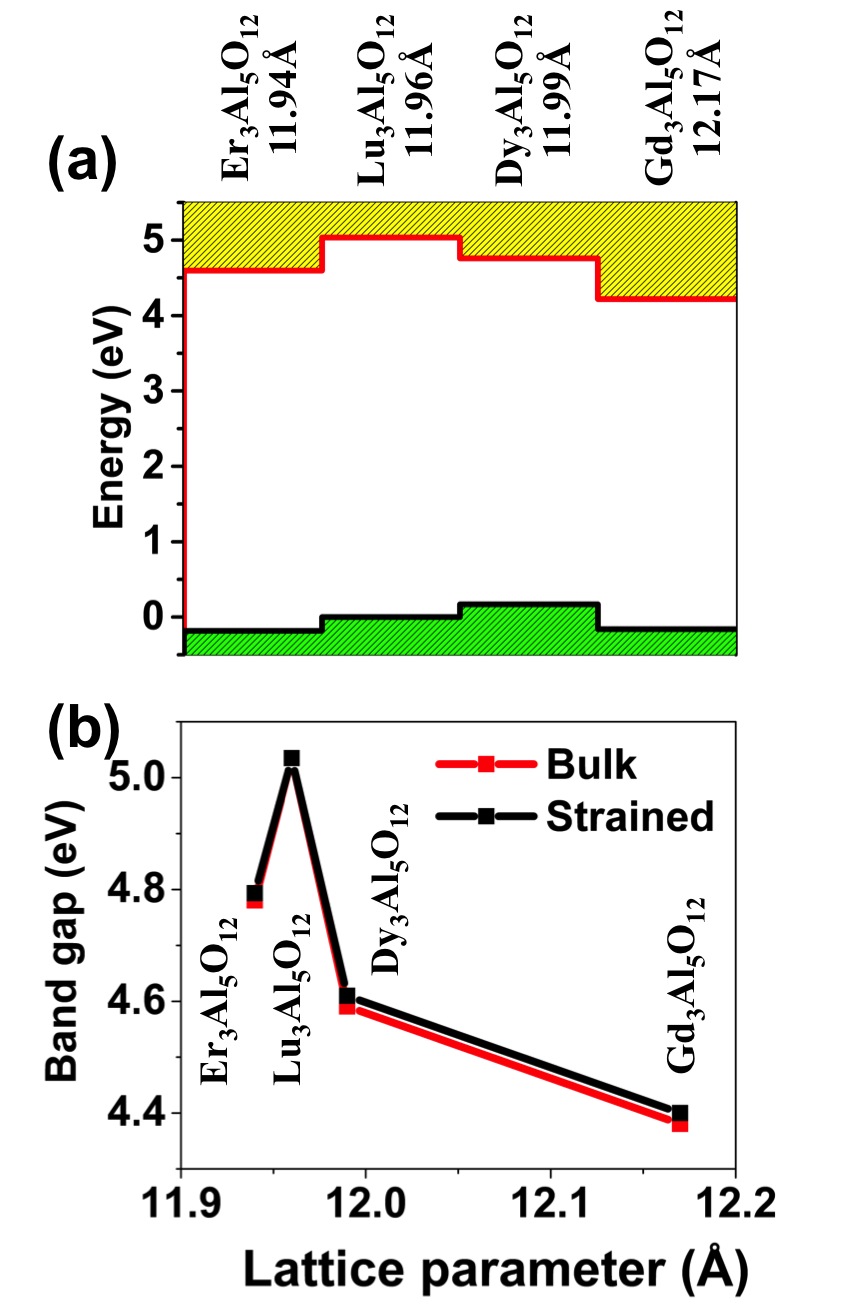}
\caption {(Color online) (a) Conduction and valence band changes for RE$_3$Al$_5$O$_{12}$ compounds, where RE = Lu, Gd, Dy and Er. (b) Change in band gap as a function of lattice parameter for the different compounds considered. The ``bulk'' (red) line indicates the band gap of the compound at its relaxed lattice constant while the ``strained'' (black) line is the band gap of the compound when placed at the lattice constant of Lu$_3$Al$_5$O$_{12}$. The band gap for the strained compound is plotted versus the compound's natural lattice constant, for ease of comparison.}
{\label{lu-substitution}}
\end{figure}

\subsection{Electronic structure variation due to Lu substitution }
Now we move to extreme case of full Lu substitution with Gd, Dy and Er to assess the effect of admixing on the RE site on the band gap and band edges of LuAG. The relative position of the band edges of RE$_3$Al$_5$O$_{12}$ (where RE = Lu, Gd, Dy, and Er) is shown in Fig.~\ref{lu-substitution}(a). Fig.~\ref{lu-substitution}(b) shows the change in the band gap as a function of lattice parameter. Substituting Lu with Gd, Dy or Er results in relatively small shifts in the band gap, which is commensurate with negligible variations in lattice parameter. In all cases the VBM and the CBM shift in the same direction, resulting in an overall band gap that is relatively constant. It is also interesting to note that Lu, Gd, Dy, and Er \textit{5d} states dominate the bottom of the conduction band.

\begin{figure*}
\includegraphics[width=7in]{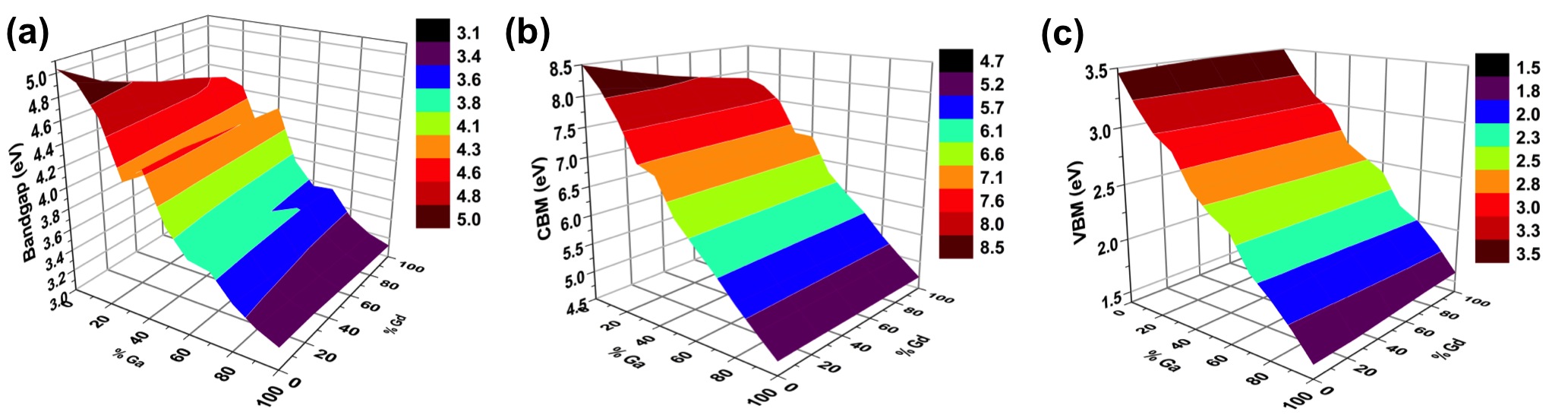}
\caption {(Color online) A contour plot of (a) band gap (eV), relative energy of the (b) conduction band minimum (eV) and (c) valence band maximum (eV) as a function of Ga substituting for Al and Gd substituting for Lu in  (Lu$_{1-x}$Gd$_x$)$_3$(Al$_{1-y}$Ga$_y$)$_5$O$_{12}$.}
{\label{concentration}}
\end{figure*}

\subsection{Effect of admixing concentration }
Finally, we have assessed the role of admixed species concentration on band structure.  That is, the above results only consider full substitution of Lu or Al cations in Lu$_3$Al$_5$O$_{12}$ rather than a partial replacement of Al or Lu, which is a more realistic scenario. Given that Gd and Ga are present in some of the multicomponent garnet compounds of interest for scintillating applications, we have systematically assessed how variations in their concentration modify the band gap and band edge position. 

Figure~\ref{concentration}(a) shows the variation in the band gap, (b) CBM and (c) VBM as a function of Ga ($x$) and Gd ($y$) concentration in (Lu$_{1-x}$Gd$_x$)$_3$(Al$_{1-y}$Ga$_y$)$_5$O$_{12}$. The garnet structure contains one crystallographically unique Lu site but two crystallographically unique Al sites, of which 40\% are octahedrally coordinated and 60\% are tetrahedrally coordinated. We used the special quasirandom structure (SQS)\cite{zunger1990special} approach to generate representative structures that mimic randomly substituted Ga and Gd amongst all of the sites. In generating the SQSs, we considered the tetrahedral and octahedral sublattices as distinct and constructed SQSs in which the cations were distributed independently on these two sublattices. These SQSs were then combined to achieve the various levels of substitutional species. This lead to situations in which all of the substitutional species were on tetrahedral sites for one composition (e.g. 30\%) and all on octahedral sites for the next composition (40\%), leading to discontinuities in the properties between those compositions. However, the change in band gap depends more significantly on the total Ga content and less on the actual distribution between tetrahedral and octahedral sites. 

It can be seen that the variation in band gap with Gd concentration is quite linear. The only deviation from linearity is observed when the position of the Ga switches from the octahedral to the tetrahedral site, as discussed above. Ga present on tetrahedral sites leads to a larger reduction in band gap and a larger CBM shift than when Ga is present on octahedral sites. Hence there is abrupt shift from 30\% to 40\%, when Ga substitution transitioned from all tetrahedral (at 30\%) to all octahedral (at 40\%)sites. In a truly random distribution of Ga, this abrupt shift would not occur and the dependence on the band gap and CBM shift on the Ga concentration would be linear throughout the composition range.

Furthermore, and as discussed above for the cases of full substitution, the change in band gap is much more sensitive to changes in the B cation than the A cation. Over most of the compositional range, the band gap is relatively insensitive to the Gd concentration, except for when the Ga content is very small. All of the change in the band gap in this compositional range is due to Ga-induced changes in the CBM. Finally, the variation of the VBM with Ga and Gd concentration is even more linear than the changes in the CBM and the overall band gap. This suggests that the changes in band edges are relatively simple functions of the dopant concentrations and that, for most of the compositional space, the effect of the two dopants is independent. 

\section{Discussion}
Our results suggest that both strain and chemistry play important roles in determining the band gap and relative position of band edges in A$_3$B$_5$O$_{12}$ garnets.  Cations with larger radii tend to produce smaller band gaps. This is accompanied by an increase in the lattice parameter. This suggests that the cation radius can be used as an initial screening parameter in the search for candidate dopants to modify the band gap. However, while we have considered extreme limits of full substitution, some of these hypothetical compounds may not be realizable experimentally.  This may explain the need to co-dope Lu$_3$Al$_5$O$_{12}$ with both Ga and Gd. Gd, having a larger radius than Lu, would help maintain the A/B radius ratio in A$_3$B$_5$O$_{12}$ garnets, stabilizing the compound. Further, also a consequence of the larger size, Gd would suppress excess antisite formation between the A and B sites as it would increase the average disparity in cation size between the two sites.

It might also be advantageous to dope or admix with smaller amounts of larger cations. For example, the band gap change for In-substituted LuAG is much larger than for Ga-substituted LuAG. One might be able to achieve the same shifts in the CBM exhibited for full substitution of Ga by relatively modest amounts of In substitution. This would provide for more opportunities for admixing strategies, as discussed below. 

To better isolate the roles of strain and chemistry, we calculated the band gap of all of the compounds considered when they are strained to the lattice constant of Lu$_3$Al$_5$O$_{12}$. These results are shown in Fig.~\ref{al-substitution}(b) and \ref{lu-substitution}(b) with the line labeled ``strained''. For both Al and Lu substitution, the band gap for the compounds at their natural lattice constant and when strained to the Lu$_3$Al$_5$O$_{12}$ lattice constant show very similar behavior. This indicates that the changes in the band edges are not simply a consequence of strain induced by changing the radii of the cations, but rather is an effect inherent in the chemistry of the cations. Thus, while the cation radius seems to correlate with the changes in band gap, it is not a direct cause of those changes.

Our results suggest admixing strategies to finely tune the band edges of complex oxide compounds for applications such as scintillators. One can imagine admixing LuAG with both Ga and As, the first to lower the CBM and remove electron traps and the second to raise the VBM and eliminated hole traps. Further, the results in Fig.~\ref{concentration} suggest a more-or-less linear relationship between the shifts in band edges and the dopant concentration. Of course, the stability of such chemically complex garnets must be examined, but by choosing the appropriate dopant species and concentrations, the band edges, in principle, can be tuned to very precise values. In fact, the results in Fig.~\ref{al-substitution} suggest that if one were to co-dope with In and either As or Sb, the band gap might be eliminated altogether. If such a compound is not thermodynamically stable, there might be other dopants that can achieve the same effect. In addition, in multicomponent garnets a positive effect as for light yield value is also expected due to local chemical composition fluctuation and related band edges fluctuation which is supposed to limit out-of-track migration of charge carriers thus supporting their immediate radiative recombination at emission centers \cite{gektin2014scintillation}.

Ce$^{3+}$ is a typical dopant used as a center for photoluminescence.  The upward shift of the VBM with admixing by As and Sb will reduce the energy gap between the VBM and the Ce$^{3+}$ ground state, which might facilitate the hole transfer from the valence band towards the Ce$^{3+}$ center in multicomponent garnet hosts. In YAG, LuAG and GGAG this energy gap has been estimated to be about 3.6 eV \cite{wu2014role}. Such a large energy gap can indeed lower the probability of fast hole transfer towards Ce$^{3+}$. An optimum gap value in this case is usually considered within 0.5-1 eV \cite{dorenbos2010fundamental}. Thus, with the right concentrations of As or Sb, this VBM-Ce gap can be reduced to optimal values. 

\section{Conclusions}
Admixing RE$_3$Al$_5$O$_{12}$ garnet compounds with Ga and Gd has led to pronounced improvements in scintillator performance, in part due to shifts in the conduction band such that the energy level of shallow defects is no longer in the forbidden gap. In this work we screen for additional admixing species using first-principles DFT, focusing on the variation of band edges in order to potentially ``band edge engineer" next-generation garnet scintillators. We have shown that certain dopants can influence the VBM or the CBM or both, which opens the door for further admixing strategies to optimize scintillator compounds. We show that substituting Al with Ga, In, As, and Sb in LuAG changes the band gap, with ionic elements (Ga and In) tend to decrease the band gap by lowering the CBM, and, on the other hand, covalent elements (In and As) tending to decrease the band gap by pushing the VBM higher in energy. In contrast, substituting Lu with Gd, Dy or Er changes neither the band gap nor the band edges to any significant degree. This study opens the possibility of tuning band gaps and band edges by admixing not only garnets but other complex oxides as well. The ability to control band gap and band edges independently is a powerful tool to optimize the performance of various materials for technological applications including not only scintillation, but also solar cells, light emitting diodes, and field effect transistors that require proper alignment of band edges across heterostructures.   

\begin{acknowledgements}
BPU was supported by the U.S. Department of Energy, Office of Science, Basic Energy Sciences, Materials Sciences and Engineering Division. MK acknowledge partial support of Czech National Science foundation grant no. P204/12/0805. Los Alamos National Laboratory, an affirmative action equal opportunity employer, is operated by Los Alamos National Security, LLC, for the National Nuclear Security Administration of the U.S. DOE under contract DE-AC52-06NA25396. 
\end{acknowledgements}

\bibliography{reference-b}

\end{document}